\documentclass[12pt,preprint]{aastex}

\shorttitle{HCO$_2^+$ in L1527} 
\shortauthors{Sakai et al.}

\usepackage{lscape}
\begin{document}

\title{Detection of HCO$_2^+$ toward the Low-Mass Protostar IRAS 04368+2557 in L1527}

\author{Nami Sakai\altaffilmark{1}, Takeshi Sakai\altaffilmark{2}, Yuri Aikawa\altaffilmark{3}, and Satoshi Yamamoto\altaffilmark{1}}

\altaffiltext{1}{Department of Physics, The University of Tokyo, Bunkyo-ku, Tokyo 113-0033, Japan}
\altaffiltext{2}{Nobeyama Radio Observatory, Minamimaki, Minamisaku, Nagano 384-1305, Japan}
\altaffiltext{3}{Department of Earth and Planetary Sciences, Kobe University, Kobe 657-8501, Japan}

\begin{abstract}
The millimeter-wave rotational emission lines ($4_{04}-3_{03}$ and $5_{05}-4_{04}$) of protonated carbon dioxide, HCO$_2^+$(HOCO$^+$), has been detected toward the low-mass class 0 protostar IRAS 04368$+$2557 in L1527 with the IRAM 30 m telescope.  This is the first detection of HCO$_2^+$ except for the Galactic Center clouds.  The column density of HCO$_2^+$ averaged over the beam size (29$^{\prime\prime}$) is determined to be $7.6 \times 10^{10}$ cm$^{-2}$, assuming the rotational temperature of 12.3 K.  The fractional abundance of gaseous CO$_2$ relative to H$_2$ is estimated from the column density of HCO$_2^+$ with an aid of a simplified chemical model.  If the HCO$_2^+$ emission only comes from the evaporation region of CO$_2$ near the protostar ($T$\raisebox{0.4ex}{$>$}\hspace{-0.75em}\raisebox{-.7ex}{$\sim$}50 K), the fractional abundance of CO$_2$ is estimated to be higher than $6.6 \times 10^{-4}$.  This is comparable to the elemental abundance of carbon in interstellar clouds, and hence, the direct evaporation of CO$_2$ from dust grain is unrealistic as a source of gaseous CO$_2$ in L1527.  A narrow line width of HCO$_2^+$ also supports this.  On the other hand, the fractional abundance of CO$_2$ is estimated to be $2.9 \times 10^{-7}$, if the source size is comparable to the beam size.  These results indicate that gaseous CO$_2$ is abundant even in the low-mass star-forming region.  Possible production mechanisms of gaseous CO$_2$ are discussed.\\
\end{abstract}


\clearpage

\section{INTRODUCTION}
Carbon dioxide (CO$_2$) is an abundant and important constituent in planetary atmospheres as well as cometary comae, and is considered to be a key species which links between interstellar chemistry and planetary chemistry.  So far solid CO$_2$ was found ubiquitously toward intermediate- to high mass star-forming regions through observations of its vibrational bands with Infrared Space Observatory (ISO) and Spitzer Space Telescope (e.g. de Graauw et al. 1996; Grakines et al. 1999; Nummelin et al. 2001; Sonnentrucker et al. 2006).  A typical abundance of solid CO$_2$ relative to H$_2$ is as high as 10$^{-5}$ to 10$^{-6}$ (Gerakines et al. 1999).  Furthermore, Dartoris et al. (2005) discovered huge amount of solid CO$_2$ even toward the low-mass class 0 protostar, L723, by observations with Spitzer Space Telescope.  Gaseous CO$_2$ was also detected in hot regions around several massive young stellar objects, where a typical abundance relative to H$_2$ was reported to be an order of 10$^{-7}$ (e.g. van Dishoeck et al. 1996; Boonman et al. 2003b; Sonnentrucker et al. 2006).  These observations clearly indicate that CO$_2$ is a major chemical component in molecular clouds.

However, the formation processes of CO$_2$ are not well understood.  In particular, a contribution of the gas phase production of CO$_2$ is quite uncertain.  Although the gas-phase chemical models predict the CO$_2$ abundance relative to H$_2$ of about 10$^{-7}$ (e.g. Lee et al. 1997), this has not been confirmed observationally in cold molecular clouds and low-mass star-forming regions.  A difficulty in measuring the abundances of gaseous CO$_2$ in these regions comes from a lack of a permanent dipole moment of CO$_2$.  Because of this, radio observations of the rotational emission lines are impossible.  Infrared observations of its vibration-rotation transitions would also be difficult because of insufficient brightness of background infrared sources or even of a lack of the sources.  An alternative way is to observe the rotational emission lines of protonated carbon dioxide, HCO$_2^+$, in the millimeter-wave region, because its abundance is directly related to the abundance of CO$_2$.  However, HCO$_2^+$ has never been detected in any cold molecular clouds and any star-forming regions except for the Galactic Center clouds like Sgr B2 and Sgr A (Thaddeus et al. 1981; DeFrees et al. 1982; Minh et al. 1991).  It has been thought that HCO$_2^+$ is produced only under a specific condition in the Galactic Center region, where shock chemistry would play an important role (e.g. Minh et al. 1988; Charnley et al. 2000).

An important breakthrough to this situation came from our preliminary line survey toward a low-mass protostar IRAS 04368+2557 in L1527 with the Nobeyama 45 m radio telescope\footnote{Nobeyama Radio Observatory is a branch of the National Astronomical Observatory of Japan, National Institutes of Natural Sciences, Japan} (NRO 45 m), by which peculiar carbon-chain chemistry in this source (Sakai et al. 2007, 2008a, 2008b) was being explored.  With a very sensitive observation, we recently found a weak emission line at the frequency of the $4_{04}-3_{03}$ line of HCO$_2^+$.  IRAS 04368+2557 is a low-mass protostar in a transient phase from class 0 to class I, which has extensively been studied by many researchers (e.g. Ohashi et al. 1997; Hogerheijde et al. 1997, 1998).  Hence, this discovery, if confirmed, is very important for understanding of the behavior of CO$_2$ during chemical evolution from molecular clouds to protoplanetary disks.  Then we carried out more sensitive observations with the IRAM 30 m telescope (IRAM 30 m) and firmly confirmed our detection of HCO$_2^+$.  In this Letter, we report the first detection of HCO$_2^+$ in the star-forming region, L1527, and discuss its astrochemical implication.\\

\section{OBSERVATIONS}
Observations of the HCO$_2^+$ lines toward IRAS 04368$+$2557 in L1527 were carried out with NRO 45 m and IRAM 30 m in 2007.  We observed the $4_{04}-3_{03}$ and the $5_{05}-4_{04}$ lines, whose frequencies are listed in Table 1.  The observed position was $(\alpha_{2000}, \delta_{2000}) = (04^{\rm h} 39^{\rm m} 53^{\rm s}.89, 26^{\circ} 03^{\prime} 11^{\prime\prime}.0)$.

In the observation with NRO 45 m, we used two SIS mixer receivers (S80 and S100) simultaneously, whose typical system temperature was about 300 K.  The main beam efficiency and the beam size of the telescope are 0.51 and 20$^{\prime\prime}$, respectively, at 85.5 GHz.  The telescope pointing was checked by observing the nearby SiO maser source (NML Tau) every hour.  The pointing accuracy was better than 8$^{\prime\prime}$.  The position-switching mode was employed for the observations, where the off position was taken at $\Delta \alpha = 30^{\prime}$, $\Delta \delta=30^{\prime}$.  The backend was an acousto-optical radiospectrometer (AOS-W), whose bandwidth is 250 MHz.  The velocity resolution is 0.87 km s$^{-1}$ at 85.5 GHz.  The intensity scale was calibrated by the chopper wheel method.

In IRAM 30 m observations, the 3 mm SIS receivers (A100 and B100) were used as front ends, whose system noise temperatures ranged from 75 K to 105 K for A100 and from 90 K to 130 K for B100.  The beam size and the main beam efficiency are 29$^{\prime\prime}$ and 0.78, respectively, at 85.5 GHz.  The telescope pointing was checked every hour by observing nearby continuum sources, and was maintained to be better than 4$^{\prime\prime}$.  The backend was an autocorrelator, VESPA.  We set the bandwidth and resolution of the individual window to be 20 MHz and 20 kHz, respectively.  The frequency resolution corresponds to the velocity resolution of 0.07 km s$^{-1}$ at 85.5 GHz.  The observation was made in the frequency switching mode with a frequency offset of 2 MHz.\\

\section{RESULTS}
The $4_{04}-3_{03}$ line of HCO$_2^+$ was first recognized toward L1527 during our preliminary line survey with NRO 45 m.  Although the line width is slightly broad due to the insufficient velocity resolution, the line was detected with the 5.0 $\sigma$ confidence level in integrated intensity (Table 1).  However, it was still necessary to confirm this detection by observing multiple rotational transitions.  Then, we made further observations with IRAM 30 m.  As a result, we successfully detected the $4_{04}-3_{03}$ and $5_{05}-4_{04}$ lines.  Although the signal-to-noise ratio of the $5_{05}-4_{04}$ line is not very good, it was detected with 4.7 $\sigma$ confidence level in integrated intensity.  The line parameters derived from Gaussian fitting are summarized in Table 1, and the line profiles are shown in Figure 1.  The $V_{\rm LSR}$ values are consistent with those of other molecules in this region (e.g. Sakai et al. 2007, 2008), no significant velocity shift being observed.  The line widths are slightly narrower than those of the other molecules ($\sim$ 0.5 km s$^{-1}$).  Because of the narrow line width and sparse spectral density, accidental matching of other lines is unlikely (cf. Sakai et al. 2008b).

The beam averaged column density of HCO$_2^+$ is determined from the observed intensities by the least-squares method.  We assume the LTE (local thermodynamic equilibrium) condition and take the effect of the optical depth into account.  Details of the procedure are described elsewhere (Sakai et al. 2008a).  The LTE assumption is justified because the density of the L1527 core ($>$10$^6$ cm$^{-3}$ for $\sim 40^{\prime\prime}$ scale) (Sakai et al. 2008a) is higher than the critical density of the observed HCO$_2^+$ transitions ($(3-6) \times 10^5$ cm$^{-3}$), where the collisional cross sections are taken from Hammami et al. (2007).  Since the upper state energies of the two observed transitions are not very different from each other, it is difficult to determine the rotational temperature of HCO$_2^+$ significantly.  Therefore, we assume the rotational temperature of 12.3 K, which is the typical value in L1527 (e.g. Sakai et al. 2008a).  The column density of HCO$_2^+$ is determined to be $7.6 \times 10^{10}$ cm$^{-2}$, where we employ the dipole moment along the a-axis of 2.85 D calculated by the B3LYP/6-311G(d, p) method (Y. Osamura, 2008, private communication).  Even when the assumed excitation temperature is changed to 30 K, the derived column density increases only by a factor of 1.8.  The optical depth is 0.006 and 0.005 for the $4_{04}-3_{03}$ and $5_{05}-4_{04}$ lines, respectively, if the beam dilution effect is not taken into account.

The source size of HCO$_2^+$ is unknown, but we could put a constraint on it from our observations.  The integrated intensity ($T_{\rm MB}$) of the $4_{04}-3_{03}$ line observed with NRO 45 m seems to be higher than that observed with IRAM 30 m in spite of a large uncertainty of the NRO 45 m data (Table 1).  This implies that the HCO$_2^+$ distribution is at least smaller than the beam size of IRAM 30 m (29$^{\prime\prime}$).  Therefore the peak column density of HCO$_2^+$ would be higher than the beam averaged value obtained above.

The present result shows the existence of HCO$_2^+$ in the low-mass star-forming core.  The column density of HCO$_2^+$ in L1527 is less than those toward Sgr B2 and Sgr A (e.g. Minh et al. 1988; Deguchi et al. 2006) by almost three orders of magnitude.  On the other hand, the HCO$_2^+$ lines were not detected in other regions including star-forming cores such as Orion KL and B335 in spite of extensive searches (Minh et al. 1988).  One possible explanation of the none-detection would be insufficient sensitivity of the previous observations.  In fact, the column density of HCO$_2^+$ in L1527 is lower than the upper limits reported by Minh et al. (1988) ($\sim 10^{12}$ cm$^{-2}$).\\

\section{DISCUSSION}
\subsection{\it Fractional Abundance of CO$_2$}
In this section, we derive the fractional abundance of the gaseous CO$_2$ with an aid of a simplified chemical model.  The abundance of HCO$_2^+$ is related to that of CO$_2$ through the following reactions;
$$ {\rm H}_3^+ + {\rm CO}_2  \rightarrow  {\rm HCO}_2^+ + {\rm H}_2:  \eqno(1) $$
$$ {\rm HCO}_2^+ + {\rm CO}  \rightarrow  {\rm HCO}^+ + {\rm CO}_2,  \eqno(2) $$
where H$_3^+$ is formed by the cosmic-ray ionization of H$_2$ and subsequent reaction of H$_2^+$ + H$_2$, and is destructed by
$$ {\rm H}_3^+ + {\rm CO}  \rightarrow  {\rm HCO}^+ + {\rm H}_2.  \eqno(3) $$
The electron recombination is less important as destruction processes of HCO$_2^+$ and H$_3^+$, and is ignored.  When we assume the steady-state condition, ratio of the number density, $n$(HCO$_2^+$)/$n$(CO$_2$), can be written as
$$ \frac{n({\rm HCO}_2^+)}{n({\rm CO}_2)} \simeq \frac{{\it k}_1 \zeta n({\rm H}_2)}{{\it k}_2{\it k}_3n({\rm CO})^2} = 1.9 \times 10^{-8} \times \frac{1}{f_{\rm CO}^{ 2} n({\rm H}_2)},  \eqno(4) $$
where $k_1$, $k_2$, and $k_3$ represent the rate coefficients for reactions (1), (2), and (3), respectively, and $\zeta$ represents the cosmic ray ionization rate, $1.3 \times 10^{-17}$ s$^{-1}$ (e.g. Aikawa et al. 2008).  Here the unit of $n$(X) is cm$^{-3}$, and the fractional abundance of CO relative to H$_2$ is denoted as $f_{\rm CO}$.  We employ the $k_1$, $k_2$, and $k_3$ values of $1.9 \times 10^{-9}$ cm$^3$ s$^{-1}$ (Burt et al. 1970), $7.8 \times 10^{-10}$ cm$^3$ s$^{-1}$ (UMIST data base), and $1.7 \times 10^{-9}$ cm$^3$ s$^{-1}$ (Kim et al. 1975), respectively.  The numerical factor in equation (4) could be larger or smaller by a factor of 3. 

It should be noted that the $n$(HCO$_2^+$)/$n$(CO$_2$) ratio is inversely proportional to the H$_2$ density, as far as $f_{\rm CO}$ is constant.  In other words, the fractional abundance of CO$_2$ relative to H$_2$ ($f_{{\rm CO}_2}$) is simply proportional to $n$(HCO$_2^+$);
$$ f_{{\rm CO}_2} \simeq 5.4 \times 10^7 \times f_{\rm CO}^{ 2} \times n({\rm HCO}_2^+).  \eqno(5) $$
If CO$_2$ exists only within the sphere with a radius of $R$ around the protostar, with the constant $f_{{\rm CO}_2}$, $n$(HCO$_2^+$) is also constant and is related to the beam averaged column density $N_{ave}$(HCO$_2^+$) as
$$ N_{ave}({\rm HCO}_2^+) = n({\rm HCO}_2^+) \times 2R \int _{0}^{(\frac{R}{R_{b}})^2}  e^{-t}\sqrt{1-(\frac{R_{b}}{R})^2t} \ dt,  \eqno(6) $$
assuming the Gaussian-shaped telescope beam with an e-fold beam radius of $R_b$.  Note that the right-hand side of equation (6) approaches to $2Rn$(HCO$_2^+$) for $R \gg R_b$ and to $n$(HCO$_2^+$)$(4 \pi R^3/3)/(\pi R_b^2)$ for $R \ll R_b$.  With equation (6), we can derive $f_{{\rm CO}_2}$ directly from the observed $N_{ave}$(HCO$_2^+$) without any assumptions on the H$_2$ density and the H$_2$ column density.  We calculate $f_{{\rm CO}_2}$ as a function of $R$, as shown in Figure 2, where we employ the $f_{\rm CO}$ value of $3.9 \times 10^{-5}$ reported by J$\o$rgensen et al. (2002).  The half-power beam radius ($R_b\sqrt{ln2}$) for IRAM 30 m is 2000 AU, where the distance to L1527 is assumed to be 140 pc.

According to Maret et al. (2004), the size of the hot region where CO$_2$ can be evaporated (\raisebox{0.4ex}{$>$}\hspace{-0.75em}\raisebox{-.7ex}{$\sim$}50 K) is estimated to be smaller than 140 AU for L1527.  If gaseous CO$_2$ traced by our HCO$_2^+$ observation are confined to this small radius, $f_{{\rm CO}_2}$ should be evaluated to be higher than $6.6 \times 10^{-4}$ by using Figure 2.  In the ice evaporation region(\raisebox{0.4ex}{$>$}\hspace{-0.75em}\raisebox{-.7ex}{$\sim$}100 K), H$_2$O and NH$_3$ would become very abundant and contribute to further destruction of HCO$_2^+$ in the innermost region.  This effect can be involved as an effectively larger $f_{\rm CO}$ value in equations (4) and (5).  Then $f_{{\rm CO}_2}$ is comparable to the elemental abundance of carbon in interstellar clouds ($7 \times 10^{-4}$).  On the other hand, Furlan et al. (2007) recently reported the mass fractional abundance of solid CO$_2$ relative to the gas in L1527 to be $1 \times 10^{-4}$ with the Spitzer observation, which corresponds to the fractional abundance of solid CO$_2$ relative to H$_2$ of $5 \times 10^{-6}$.  This is much lower than the $f_{{\rm CO}_2}$ value derived from our observation.  Hence, the direct evaporation of CO$_2$ is very unrealistic as a major source of gaseous CO$_2$ traced by HCO$_2^+$ in L1527.  This would further be supported by the narrow line width (0.57 km s$^{-1}$) of HCO$_2^+$ in comparison with the line width of the very high excitation line of HC$_5$N ($J=32-31$, 85.2 GHz, $E_u$=67 K) observed with IRAM 30 m in L1527 (Sakai et al. in preparation).

If we assume $f_{{\rm CO}_2}$ of $\sim 5 \times 10^{-7}$, which is close to the value reported in Orion-KL IRc2 by the infrared observation (Boonman et al. 2003a), $R$ is derived to be 1600 AU.  On the other hand, $f_{{\rm CO}_2}$ is still as high as $2.9 \times 10^{-7}$, even if HCO$_2^+$ is extended over the beam size of IRAM 30 m.  Therefore, substantial amount of CO$_2$ exists even in the low temperature region where thermal evaporation of CO$_2$ from grain mantles is very inefficient.  It should be noted that the interaction between outflows and ambient gas would not be responsible for the CO$_2$ production in L1527, because the line width of HCO$_2^+$ is narrow and no velocity shift is observed.

\subsection{\it Origin of Gaseous CO$_2$}
As described above, most of gaseous CO$_2$ in L1527 does not originate from direct evaporation of solid CO$_2$, but from gas-phase reactions.  Since the gas phase production of CO$_2$ is not well understood observationally (e.g. van Dishoeck et al. 2004), we discuss a few possibilities.  First we compare our result with the chemical model calculation by Lee et al. (1997).  According to their 'new standard' model, the gas phase abundance of CO$_2$ is expected to be $2.8 \times 10^{-7}$ and $1.4 \times 10^{-7}$ at $10^{5.5}$ yr and the steady state, respectively, for the H$_2$ density of 10$^5$ cm$^{-3}$.  This is almost consistent with our result in L1527, if the CO$_2$ distribution is extended over the beam size of IRAM 30 m.

Secondly, CO$_2$ might be produced in a hot region through a reaction of OH and CO.  Since this reaction has activation barrier of 176 K (UMIST data base; See also Talbi and Herbst 2002 for the potential energy surface.), it is efficient in a small region near the protostar, like a hot core or a hot corino.  In such a hot region, the abundance of H$_2$CO is enhanced due to mantle evaporation, which would supply HCO through the electron recombination reaction of H$_2$COH$^+$ formed from H$_2$CO and H$_3^+$.  Then, CO$_2$ will be produced through the HCO $+$ O reaction (e.g. Snyder et al. 1985).  However, these high temperature mechanisms occur in a much smaller region than the CO$_2$ evaporation region, and hence, they can not be major contributors.

Since L1527 shows warm carbon-chain chemistry (WCCC) (Sakai et al. 2008a), we finally consider a possible CO$_2$ formation related to the WCCC.  In the WCCC, CH$_4$ evaporated from grain mantles drives the carbon-chain chemistry.  At the same time, HCO would be produced by various reactions, which successively forms CO$_2$ by the reaction with the atomic oxygen.  The warm region, where CH$_4$ can be evaporated, is substantially larger than the region where CO$_2$ can be evaporated, and hence, CO$_2$ would be distributed over the size of WCCC (Sakai et al. 2008a).  Rather compact distribution of HCO$_2^+$ inferred from the beam dilution effect, if correct, favors this scenario rather than the production of CO$_2$ in a cold cloud.  In this relation, the HCO ($1_{01}-0_{00}$, 86.7 GHz) line is detected toward L1527 in our preliminary line survey, which may also support the above idea.  Furthermore, an enhancement of gaseous CO$_2$ in the warm region, where CH$_4$ is evaporated, is predicted in a chemical model of a dynamically evolving cloud by Aikawa et al. (2008, Figs. 6 and 10).  In order to establish the formation pathway of CO$_2$, distribution of HCO$_2^+$ in L1527 is a key.  A high spatial resolution observation with an interferometer would be necessary.  By comparing the result with the detailed chemical model calculations, we will be able to investigate the contributions of cold gas phase chemistry and WCCC to the formation of CO$_2$.  Sensitive observations of HCO$_2^+$ toward various clouds would also be needed to understand the origin of gaseous CO$_2$.

\acknowledgments
We are grateful to Eric Herbst for his invaluable discussion.  We also thank to Yoshihiro Osamura for communicating the results of his ab intio calculations.  We thank to the staff of the IRAM 30 m telescope and the NRO 45 m telescope for their excellent support.  This study is supported by Grant-in-Aids from Ministry of Education, Culture, Sports, Science, and Technologies (14204013, 15071201, and 19-6825).\\

\newcommand{\ana}{A\&A}
\newcommand{\nt}{Nature}
\newcommand{\mn}{MNRAS}
\newcommand{\rsi}{RSI}
\newcommand{\jms}{JMS}






\begin{landscape}
\begin{table}[htbp]
\caption{Observed Line Parameters for HCO$_2^+$}
\begin{center}
\begin{tabular}{lcccccc}\hline\hline
Transition&Frequency&$T_{\rm MB}^{\rm a}$&$dv^{\rm a}$&$V_{\rm LSR}^{\rm a}$&rms$^{\rm b}$&$\int T_{\rm MB} dv$($3\sigma$)\\
$\quad$&[GHz]&[K]&[km s$^{-1}$]&[km s$^{-1}$]&[mK]&[K km s$^{-1}$] \\ \hline
\multicolumn{7}{c}{IRAM 30 m} \\ \hline
$4_{04}-3_{03}$&85.5315123&0.058(8)&0.32(5)&5.88(2)&4.5&0.020(4) \\
$5_{05}-4_{04}$$^{\rm c}$&106.9135639&0.044(6)&0.42(6)&5.88(3)&10.1&0.020(13) \\ \hline
\multicolumn{7}{c}{NRO 45 m$^d$} \\ \hline
$4_{04}-3_{03}$&85.5315123&0.039(11)&1.0(3)&6.2(2)&8.0&0.040(24) \\
\hline
\end{tabular}
\end{center}

Note.--- The numbers in parentheses represent the errors in units of the last significant digits. \\
$^{\rm a}$ Obtained by the Gaussian fit.\\
$^{\rm b}$ The rms noise averaged over the linewidth.\\
$^{\rm c}$ Five successive channels of VESPA are bound to improve the S/N ratio.\\
$^{\rm d}$ The line width becomes broad due to the insufficient velocity resolution of AOS-W.
\label{tab:tbl-1}
\end{table}
\end{landscape}
\clearpage

\begin{figure}
\epsscale{0.5}
\plotone{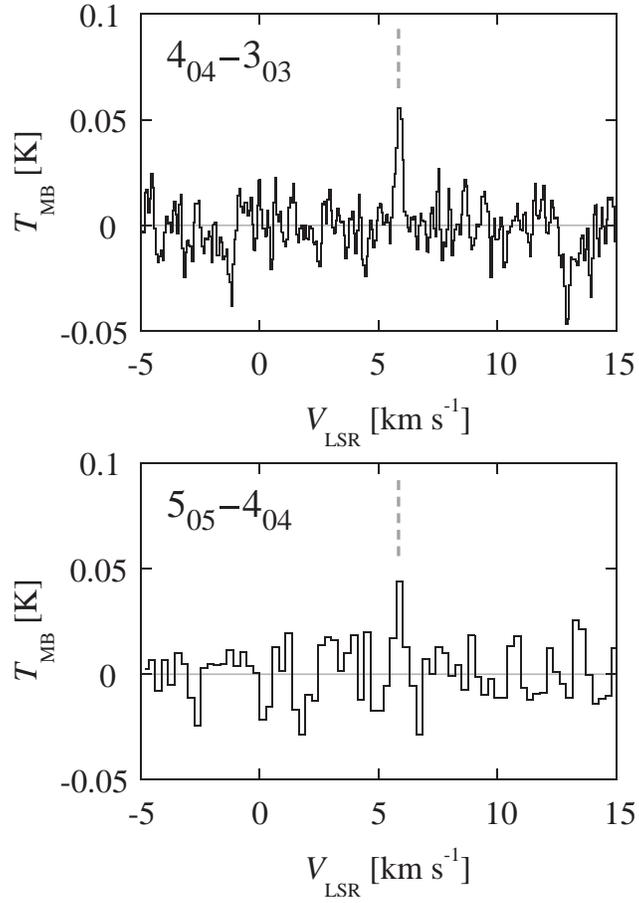}
\caption{Spectral line profiles of HCO$_2^+$ ($4_{04}-3_{03}$ and $5_{05}-4_{04}$) observed toward L1527.  Negative features at -1 and 13 km s$^{-1}$ in the $4_{04}-3_{03}$ spectrum are frequency-switching artifacts.}
\label{fig:f1}
\end{figure}
\clearpage

\begin{figure}
\epsscale{0.5}
\plotone{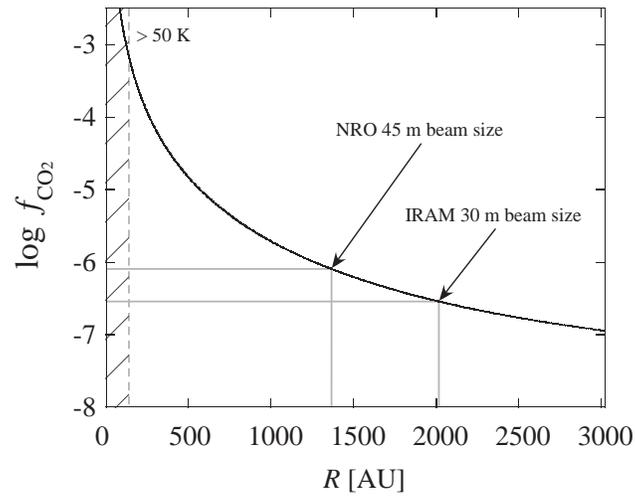}
\caption{Fractional abundance of CO$_2$ relative to H$_2$ as a function of the assumed radius of the emitting region of the HCO$_2^+$ line.}
\label{fig:f2}
\end{figure}


\begin{thebibliography}{}

\bibitem[Aikawa et al. (2008)]{ai08} Aikawa, Y., Wakelam, V., Garrod, R. T., and Herbst, E. 2007, \apj, in press.
\bibitem[Boonman et al. (2003a)]{bo03a} Boonman, A. M. S., van Dishoeck, E. F., Lahuis, F., Doty, S. D., Wright, C. M., and Rosenthal, D. 2003a, \ana, 399, 1047.
\bibitem[Boonman et al. (2003b)]{bo03b} Boonman, A. M. S., van Dishoeck, E. F., Lahuis, F., and Doty, S. D. 2003b, \ana, 399, 1063.
\bibitem[Burt et al. (1970)]{bu70} Burt, J. A., Dunn, J. L., McEwan, M. J., Sutton, M. M., Roche, A. E., and Schiff, H. I. 1970, J. Chem. Phys., 52, 6062.
\bibitem[Charnley et al. (2000)]{ch00} Charnley, S. B. and Kaufman, M. J. 2000, \apj, 529, L111.
\bibitem[Dartois et al. (2005)]{da05} Dartois, E., Pontoppidan, K., Thi, W. -F., and Mu$\tilde{\rm n}$oz Caro, G. M. 2005, \ana, 444, L57.
\bibitem[DeFrees et al. (1982)]{de82} DeFrees, D. J., Loew, G. H., and McLean, A. D., 1982, \apj, 254, 405.
\bibitem[de Graauw et al. (1996)]{de96} de Graauw, Th. et al. 1996, \ana, 315, L345.
\bibitem[Deguchi et al. (2006)]{de06} Deguchi, S., Miyazaki, A., and Minh, Y. C. 2006, PASJ, 58, 979.
\bibitem[Furlan, et al. (2007)]{fu07} Furlan, E. et al. 2007, ApJS, in press.
\bibitem[Gerakines et al. (1999)]{ge99} Gerakines, P. A. et al. 1999, \apj, 522, 357.
\bibitem[Hammami et al. (2007)]{ha07} Hammami, K., Lique, F., Jaidane, N., Ben Lakhdar, Z., Spielfiedel, A., and Feautrier, N. 2007, \ana, 462, 789.
\bibitem[Hogerheijde et al. (1997)]{ho97} Hogerheijde, M. R., van Dishoeck, E. F., Blake, G. A., and van Langevelde, H. J. 1997, \apj, 489, 293.
\bibitem[Hogerheijde et al. (1998)]{ho98} Hogerheijde, M. R., van Dishoeck, E. F., Blake, G. A., and van Langevelde, H. J. 1998, \apj, 502, 315.
\bibitem[J$\o$rgensen et al. (2002)]{jo02} J$\o$rgensen, J. K., Sch$\ddot{\rm o}$ier, F. L., and van Dishoeck, E. F. 2002, \ana, 389, 908.
\bibitem[Kim et al. (1975)]{ki75} Kim, J. K., Theard, L. P., and Huntress, W. T. 1975, Chem. Phys. Lett., 32, 610.
\bibitem[Lee et al. (1996)]{le96} Lee, H. -H., Bettens, R. P. A., and Herbst, E. 1996, A\&AS, 1996, 119, 111.
\bibitem[Maret et al. (2004)]{ma04} Maret, S. et al. 2004, \ana, 416, 577.
\bibitem[Minh et al. (1988)]{mi88} Minh, Y. C., Irvine, W. M., and Ziurys, L. M. 1988, \apj, 334, 175.
\bibitem[Minh et al. (1991)]{mi91} Minh, Y. C., Brewer, M. K., Irvine, W. M., Friberg, P., and Johansson, L. E. B. 1991, \ana, 244, 470.
\bibitem[Nummelin et al. (2001)]{nu01} Nummelin, A., Whittet, D. C. B., Gibb, E. L., Gerakines, P. A., and Chiar, J. E. 2001, \apj, 558, 185.
\bibitem[Ohashi et al. (1997)]{oh97} Ohashi, N., Hayashi, M., Ho, P. T. P., and Momose, M. 1997, \apj, 475, 211.
\bibitem[Sakai et al. (2007)]{sa07} Sakai, N., Sakai, T., Osamura, Y., and Yamamoto, S. 2007, \apj, 667, L65.
\bibitem[Sakai et al. (2008a)]{sa08a} Sakai, N., Sakai, T., Hirota, T., and Yamamoto, S. 2008a, \apj, 672, 371.
\bibitem[Sakai et al. (2008b)]{sa08b} Sakai, N., Sakai, T., and Yamamoto, S. 2008b, \apj, 673, L71.
\bibitem[Talbi et al. (2002)]{ta02} Talbi, D. and Herbst, E. 2002, \ana, 386, 1139.
\bibitem[Thaddeus et al. (1981)]{th81} Thaddeus, P., Gu$\rm{\acute{e}}$lin, M., and Linke, R. A. 1981, \apj, 246, L41.
\bibitem[van Dishoeck et al. (1996)]{va96} van Dishoeck, E. F. et al. 1996, \ana, 315, L349.
\bibitem[van Dishoeck et al. (2004)]{va04} van Dishoeck, E. F. 2004, Annu. Rev. Astron. Astrophys, 42, 119.



\end{thebibliography}
\end{document}